\documentclass{article}

\usepackage{graphicx}
\usepackage{amssymb}

\newcommand{\stMIN}{{\scriptscriptstyle{}(-)}}
\newcommand{\stPLU}{{\scriptscriptstyle{}(+)}}
\newcommand{\stPLUMIN}{{\scriptscriptstyle{}(\pm)}}
\newcommand{\stMINPLU}{{\scriptscriptstyle{}(\mp)}}
\newcommand{\stSHE}{{\scriptscriptstyle{}(3)}}
\newcommand{\proof}[1]{\noindent{\small\textit{Proof:} #1 $\square$\par}}
\newtheorem{prop}{Proposition}
\newcommand{\bm}[1]{\mbox{\textbf{\textit{#1}}}}

\title{Vacuum and semiclassical gravity:\\
a difficulty and its bewildering significance\footnote{Submitted to the
Proceedings of Science (\texttt{http://pos.sissa.it}.)}}

\author{Stefano Ansoldi\footnote{{}Email: \texttt{ansoldi@trieste.infn.it};
Webpage: \texttt{http://www-dft.ts.infn.it/$\sim$ansoldi};
Mailing address: Dipartimento di Matematica e Informatica,
Universit\`{a} degli Studi di Udine, via delle Scienze 206, I-33100 Udine (UD), ITALY.}\\
{\small\it{}Department of Physics, University of Kyoto, Kyoto, JAPAN and}\\
{\small\it{}International Center for Relativistic Astrophysics (I.C.R.A), Pescara, ITALY and}\\
{\small\it{}Istituto Nazionale di Fisica Nucleare (I.N.F.N.), Sezione di Trieste, ITALY}}

\date{\vspace*{1cm}{\small{}Talk given at the workshop\\
\emph{From Quantum to Emergent Gravity: Theory and Phenomenology,}\\
June, 11-15 2007,\\
International School for Advanced Studies (SISSA/ISAS),\\
Trieste, Italy\\[-1mm]
\rule{9.3cm}{0.5pt}\\[-1mm]
 Conference Webpage: \texttt{http://www.sissa.it/app/QGconference}}\\[-2mm]
\rule{9.3cm}{0.5pt}}

\begin{document}

\maketitle

\begin{abstract}
\begin{center}
{We review a long-standing difficulty in some semiclassical
models of vacuum and vacuum decay. Surprisingly enough these models, careless of
their transparent formulation, are affected by, both, technical and conceptual
issues. After proving some general results that are relevant for, both,
the Euclidean and Lorentzian sectors of their dynamics, we briefly
highlight their importance in connection with the issues discussed before,
arguing that their solution might be interesting in our quest for
quantum gravity.\\[1.5cm]
\rule{12.5cm}{0.5pt}\\
\mbox{\footnotesize{}Talk online at
http://www.sissa.it/app/QGconference/TALKS/wednesday/ansoldi/index.html}}\\[-1mm]
\rule{12.5cm}{0.5pt}
\end{center}
\end{abstract}

\maketitle

\clearpage

\section{Introduction}

The study of vacuum properties and decay is a fascinating subject, and it becomes even more interesting
when its interplay with gravitation is considered. A natural context to study properties
of vacuum is early universe cosmology,
where vacuum energy density plays a dominant role. For this reason, early
after the first studies of vacuum and vacuum decay
\cite{bib:VacuumDecay}, gravity entered the scene
\cite{bib:PhReD1980..21..3305L,bib:Bubbles01,bib:PhReD1987..35..1747G,bib:Bubbles02}:
in most models thin relativistic shells are used to describe the dynamics of vacuum bubbles.
Classical models already reveal interesting properties, but they tend to be affected by
issues, in connection with stability and the presence of singularities.
To avoid these problems, quantum models (mostly semiclassical) have been
developed \cite{bib:NuPhy1990B339...417G,bib:QuantumModels}:
vacuum decay is then described in the WKB approximation as spacetime tunnelling
driven by the thin shell. Despite the earliest of these works date back to more than 30
years ago, issues left open in the original formulation \cite{bib:NuPhy1990B339...417G}
have not yet been solved \cite{bib:NuPhy1990B339...417G,bib:QuantumTroubles}.
Here we review some of these issues from a general perspective. We also point out that,
although the existing literature mostly focuses on applications relevant for
early universe cosmology, the open problems are of a more general nature and can be
recognized as a general difficulty in the semiclassical quantization of the (general relativistic)
shell system. After a preliminary review of some background material in Sec.$\:$\ref{sec:predef},
we prove in Sec.$\:$\ref{sec:genres} a collection of general properties of shell dynamics in
spherically symmetric, but otherwise arbitrary, configurations.
Then, in Sec.$\:$\ref{sec:tunpro}, we present in the perspective of
these results some issues related with the WKB description of
the tunnelling process; a concluding discussion follows in Sec.$\:$\ref{sec:dissec}.
Apart from the included bibliography, additional references
can be also found in \cite{bib:ClQuG2002..19..6321A,bib:ExtraBiblio}.

\section{\label{sec:predef}Background, conventions and notations}

\begin{figure}[htb]
\begin{center}
\begin{tabular}{|c|c|}
\hline
\includegraphics[width=7cm]{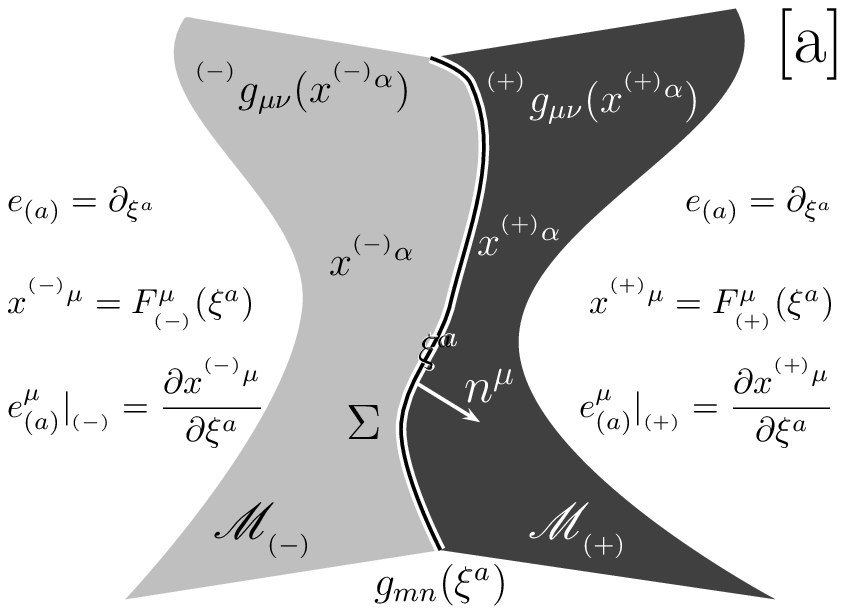}
&
\includegraphics[width=7cm]{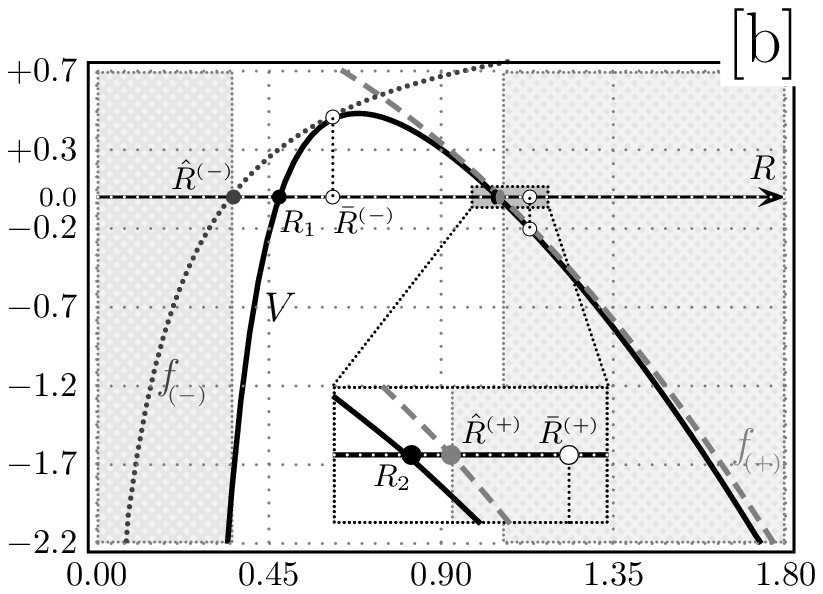}
\\
\hline
\end{tabular}
\caption{\label{fig:geoquadef}{\footnotesize{}Definition of the geometrical quantities
(panel [a]) which are important
to characterize a \emph{junction} according to the notations and conventions defined
in Sec.$\:${\protect\ref{sec:predef}} on page {\protect\pageref{sec:predef}}. In the spherically symmetric case
the dynamics of the system can be reduced to a one-dimensional effective classical problem
\cite{bib:PhReD1987..35..1747G,bib:PhReD1989..40..2511S,bib:NuPhy1990B339...417G,bib:ClQuG1997..14..2727S}.
All the properties
of the solutions can be qualitatively obtained from diagrams like the one in panel [b]: turning points are $R _{1}$
and $R _{2}$ (see also the zoomed in area), zeroes of $f _{\stPLUMIN}$ are $\hat{R} ^{\stPLUMIN}$, points at
which the signs $\epsilon _{\stPLUMIN}$ vanish are $\bar{R} ^{\stPLUMIN}$ (where $f _{\stPLUMIN}$ are tangent to $V (R)$).
Three trajectories are possible, two classical ones (the \emph{bounded} $[0 , \hat{R} _{1}]$ and the \emph{unbounded} or
\emph{bounce} $[ \hat{R} _{2} , + \infty)$) and a tunnelling one, $[ \hat{R} _{1} , \hat{R} _{2} ]$. The two grayed areas
highlight a $T ^{\stMIN}$ region (on the left) and a $T ^{\stPLU}$ region (on the right), respectively.
As discussed in
the text, the vanishing of $\epsilon _{\stPLU}$ at $\bar{R} ^{\stPLU}$ indicates a difficulty in the description
of the tunnelling process.}}
\rule{\textwidth}{0.5pt}
\end{center}
\end{figure}

Throughout the paper curvature conventions follow
\cite{bib:Freem1970...1..1279W}, with Greek indices taking the values $0,1,2,3$, lowercase Latin
indices taking the values $1, 2, 3$, and uppercase latin indices the values $0, 1$. We
consider two parts ${\mathcal{M}} _{\stMIN}$ and ${\mathcal{M}} _{\stPLU}$ of two spacetimes and assume
that they have a common timelike part $\Sigma$ in their boundaries. Various quantities related to the
submanifolds ${\mathcal{M}} _{\stMIN}$, ${\mathcal{M}} _{\stPLU}$ and $\Sigma$ are defined
in the table below and in Fig.$\:$\ref{fig:geoquadef}, panel [a].
\begin{center}
\begin{tabular}{||c||c|c|c|c|c|c|c||}
\hline
{\footnotesize{}Submanifold} &
{\footnotesize{}Dim.} &
{\footnotesize{Coordinate}} &
{\footnotesize{}Holonomic} &
{\footnotesize{}Metric} &
{\footnotesize{}Signature} &
{\footnotesize{}Covariant} &
{\footnotesize{}Matter}
\\[-2mm]
{\footnotesize{}} &
{\footnotesize{}} &
{\footnotesize{}system} &
{\footnotesize{}basis} &
{\footnotesize{}components} &
{\footnotesize{}} &
{\footnotesize{}Derivative} &
{\footnotesize{}Content}
\\
\hline & & & & & & &\\[-5mm]
\hline
${\mathcal{M}} _{\stMIN}$ &
$4$ &
$x ^{\stMIN\alpha}$ &
$\partial ^{\stMIN} _{\mu}$ &
${}^{\stMIN}\!g _{\mu \nu} (x ^{\stMIN\alpha})$ &
$2$ &
${}^{\stMIN}\!\nabla _{\mu}$, $(\dots) ^{\stMIN} _{; \mu}$ &
${}^{\stMIN}\!T _{\mu \nu}$ \\
\hline
${\mathcal{M}} _{\stPLU}$ &
$4$ &
$x ^{\stPLU\alpha}$ &
$\partial ^{\stPLU} _{\mu}$ &
${}^{\stPLU}\!g _{\mu \nu} (x ^{\stPLU\alpha})$ &
$2$ &
${}^{\stPLU}\!\nabla _{\mu}$, $(\dots) ^{\stPLU} _{; \mu}$ &
${}^{\stPLU}\!T _{\mu \nu}$ \\
\hline
$\Sigma$ & $3$ &
$\xi ^{a}$ &
$\bm{e} _{(a)} = \partial _{\xi ^{a}}$ &
$g _{m n} (\xi ^{a})$ &
$1$ &
${}^{\stSHE}\!\nabla _{a}$, $(\dots) _{| a}$ &
$S _{m n}$ \\
\hline
\end{tabular}
\end{center}
\noindent{}In the above setup, we \emph{locally} define the embedding of $\Sigma$
in ${\mathcal{M}} _{\stPLUMIN}$ as $x ^{\stPLUMIN\mu} = F ^{\mu} _{\stPLUMIN} (\xi ^{a})$. Moreover we assume
$
{}^{\stMIN}\!g _{\mu \nu} (x ^{\stMIN\rho})
e ^{\mu} _{(m)}
e ^{\nu} _{(n)} | _{x ^{\rho} = F _{\stMIN} ^{\rho} (\xi ^{a})} =
{}^{\stPLU}\!g _{\mu \nu} (x ^{\stPLU\sigma}) e ^{\mu} _{(m)} e ^{\nu} _{(n)} | _{x ^{\sigma} = F _{\stPLU} ^{\sigma} (\xi ^{a})} =
g _{m n} (\xi ^{a}),
$
so that the metric on $\Sigma$ is well defined (by assumption, it is also non degenerate).
Various geometric quantities, as for instance the normal\footnote{We assume that
it points from ${\mathcal{M}} _{\stMIN}$ to ${\mathcal{M}} _{\stPLU}$; moreover, since $\Sigma$ is
timelike, \emph{the normal is also transverse}.} to $\Sigma$, $\bm{n}$, or the extrinsic curvature of $\Sigma$,
$K _{ij} \stackrel{\mathrm{def.}}{=} - n _{\alpha} e _{(j)} ^{\beta} \nabla _{\beta} e _{(i)} ^{\alpha}$,
are calculated with respect to the embedding of $\Sigma$ in ${\mathcal{M}} _{\stMIN}$ or
${\mathcal{M}} _{\stPLU}$. We specify this using ``${\scriptstyle(\pm)}$'' as a sub/superscript or using ``$| _{\stPLUMIN}$''.
The \emph{jump} of quantities across $\Sigma$, is shortly denoted as
``$[(\dots{})]$''; consistently, square brackets \emph{are not} used with any other meaning.

The dynamics of $\Sigma$ and of its energy matter content in the spacetime manifold
$\overline{{\mathcal{M}} _{\stMIN} \cup {\mathcal{M}} _{\stPLU}}$ (the overbar denotes the closure)
is determined by Israel junction conditions
\cite{bib:IsraelJunctions},
which relate the jump of the extrinsic curvature of $\Sigma$ to the surface stress-energy tensor
$S _{mn}$ and its trace $S$:
\begin{equation}
    [K _{ij}] \equiv K _{ij} ^{-} - K _{ij} ^{+}
    =
    8 \pi \left( S _{ij} - \frac{g _{ij}}{2} S \right)
    .
\label{eq:isrjuncon}
\end{equation}
In particular, we are interested in the spherically symmetric reduction of the above equations,
i.e. we consider the special case in which $\Sigma$ describes the spherically symmetric evolution
of a sphere. A general spherically symmetric four dimensional spacetime
can always be considered as a product of a two sphere ${\mathbb{S}} ^{2}$ and of another,
again two dimensional, Lorentzian manifold ${\mathbb{M}} _{2}$. Therefore, the metric
${}^{(4)}\!\bm{g}$ on the four dimensional manifold can always be decomposed in the form
\[
    {}^{(4)}\!\bm{g} = \left( \matrix{\bm{g} _{{\mathbb{M}} _{2}} & {\mathbb{O}} \cr {\mathbb{O}} & \bm{g} _{{\mathbb{S}} ^{2}}} \right) , \quad
    \bm{g} _{{\mathbb{M}} _{2}} = \left( \matrix{ \gamma _{11} (x _{1} , x _{2}) &  \gamma _{12} (x _{1} , x _{2}) \cr  \gamma _{12} (x _{1} , x _{2}) &  \gamma _{22} (x _{1} , x _{2}) } \right) , \quad
    \bm{g} _{{\mathbb{S}} ^{2}} = \rho ^ {2} ( x _{1} , x _{2} ) \left( \matrix{ 1 & 0 \cr 0 & \sin ^{2} \Theta} \right) ,
\]
where we denote with $(x _{1} , x _{2})$ coordinates in ${\mathbb{M}} _{2}$ and with $(\Theta , \Phi)$ the usual
coordinates in ${\mathbb{S}} ^{2}$. Thanks to general covariance, we can subject
$\bm{g} _{{\mathbb{M}} _{2}}$ to two conditions so that the four dimensional metric
can be locally described by only two functions: a natural choice is $\rho (x _{1} , x _{2})$
plus another function coming from ${\mathbb{M}} _{2}$, which can be wisely chosen to be the
invariant $\Delta = \gamma ^{AB} \rho _{,A} \rho _{,B}$ (the modulus square of the vector normal to
the $\rho (x _{1}, x _{2}) = \mathrm{const.}$ surfaces). If $\Delta > 0$ we are in what is called
an $R$-region, whereas if $\Delta < 0$ we are in a $T$-region
\cite{bib:TRRegions}. When we apply the above decomposition in
${\mathcal{M}} _{\stPLUMIN}$ we can conveniently use $\Delta _{\stPLUMIN}$ to describe various properties
of the dynamics of the shell in a coordinate invariant way
\cite{bib:ClQuG2001..18..2195O}.

In the following it is also useful
to further specialize ${}^{\stPLUMIN} g _{\mu \nu}$ and $g _{m n}$, defined before, as follows.
\begin{center}
\begin{tabular}{||c||c|c|c||}
\hline
{\footnotesize{}Manifold} &
{\footnotesize{Coordinate system}} &
{\footnotesize{}Metric components} &
{\footnotesize{}Matter content}
\\
\hline & & &\\[-5mm]
\hline
${\mathcal{M}} _{\stMIN}$ &
$(t _{\stMIN} , \theta , \phi , r _{\stMIN})$ &
$\mathrm{diag}( - f _{\stMIN} (r _{\stMIN}) , r _{\stMIN} ^{2} , r  _{\stMIN} ^{2} \sin ^{2} \theta , 1 / f _{\stMIN} (r  _{\stMIN}) )$ &
spherically symm. \\
\hline
${\mathcal{M}} _{\stPLU}$ &
$(t _{\stPLU} , \theta , \phi , r _{\stPLU})$ &
$\mathrm{diag}( - f _{\stPLU} (r _{\stPLU}) , r _{\stPLU} ^{2} , r  _{\stPLU} ^{2} \sin ^{2} \theta , 1 / f _{\stPLU} (r  _{\stPLU}) )$ &
spherically symm. \\
\hline
$\Sigma$ &
$(\tau , \theta , \phi)$ &
$\mathrm{diag}(-1 , R (\tau) ^{2} , R  (\tau) ^{2} \sin ^{2} \theta$ ) &
$M (R)$ \\
\hline
\end{tabular}
\end{center}
\noindent{}We now see that $\mathrm{sign} ( \Delta _{\stPLUMIN}) = \mathrm{sign} ( f _{\stPLUMIN})$, so that
if $f _{\stPLUMIN} < 0$ we are in a $T ^{\stPLUMIN}$-region of ${\mathcal{M}} _{\stPLUMIN}$,
whereas if $f _{\stPLUMIN} > 0$ we are in
an $R ^{\stPLUMIN}$-region of ${\mathcal{M}} _{\stPLUMIN}$. Moreover,
after using all the freedom to fix the coordinate systems, $R (\tau)$ is the only degree of freedom
($\tau$ is the proper time of an observer comoving with $\Sigma$) describing the dynamics of the shell
and only one junction condition is non-trivial:
\begin{equation}
    R
    \left[ \epsilon \sqrt{\dot{R} ^{2} + f (R)} \right]
    =
    M (R)
    \quad \mathrm{where} \quad
    \left . n ^{\alpha} \partial _{\alpha} r _{\stPLUMIN} \right | _{\Sigma}
    =
    \epsilon _{\stPLUMIN} \sqrt{ \dot{R} ^{2} + f _{\stPLUMIN} (R)}
    ;
\label{eq:juncon}
\end{equation}
an overdot denotes a derivative with respect to $\tau$,
$\epsilon _{\stPLUMIN}$ are signs (we will come back later to their meaning)
and $M (R)$ describes the matter energy content of the shell, after its
equation of state has been specified. The only nontrivial junction condition (\ref{eq:juncon})
can then be rewritten in the form
\begin{equation}
    \dot{R} ^{2} + V (R) = 0
    \quad \mathrm{where} \quad
    V (R) = - \frac{(R ^{2} f _{\stMIN} (R) + R ^{2} f _{\stPLU} (R) - M ^{2} (R)) ^{2} - 4 R ^{4} f _{\stMIN} (R) f _{\stPLU} (R)}{4 M ^{2} (R) R ^{2}}
    .
\label{eq:claeffequ}
\end{equation}
A closed form expression for the $\epsilon _{\stPLUMIN}$ signs not involving $\dot{R}$ can also be obtained starting
from (\ref{eq:juncon}):
\begin{equation}
    \epsilon _{\stPLUMIN}
    =
    \mathrm{sign} \{ M (R) ( R ^{2} f _{\stMIN} (R) - R ^{2} f _{\stPLU} (R) \mp M ^{2} (R) ) \}
    .
\label{eq:epssig}
\end{equation}
Eqs.$\:$(\ref{eq:claeffequ}) and (\ref{eq:epssig}) form a set of equations which is
completely equivalent to (\ref{eq:juncon}) (see also Prop.$\:$\ref{prop:radno_res}
below). \emph{Classical solutions} of the junction condition can exist only in the region $V (R) \leq 0$
and their turning points are solutions of the equation $V (R) = 0$. We call
\emph{tunnelling trajectories} the solutions in the inverted potential $- V (R)$.
Also notice, that the junction condition (\ref{eq:isrjuncon}) is a first order equation,
(it contains $\dot{R}$ but not $\ddot{R}$) and it is not the \emph{equation of motion}
of the system but a first integral of it \cite{bib:ClQuG1997..14..2727S,bib:LagFormShells}.
The second order equation of motion can be obtained
from an effective Lagrangian, $L _{\mathrm{EFF}}$, that describes
the dynamics of the \emph{only} remaining degree of freedom $R (\tau)$.
If
\begin{equation}
    H _{\mathrm{EFF}} (R , \dot{R})
    =
    R
    \left[
        \epsilon \sqrt{\dot{R} ^{2} + f}
    \right]
    -
    M (R)
    \: \: \: \: \: \: \mathrm{and} \: \: \: \: \: \:
    P _{\mathrm{EFF}} (R , \dot{R})
    =
    R
    \left[
        \tanh ^{-1}
        \left(
            \frac{\dot{R}}{\epsilon \sqrt{\dot{R} ^{2} + f}}
        \right) ^{\mathrm{sgn} (f)}
    \right]
\label{eq:effmom}
\end{equation}
are the effective (super)hamiltonian and effective momentum, respectively,
we have that the $L _{\mathrm{EFF}} = P _{\mathrm{EFF}} \dot{R} - H _{\mathrm{EFF}}$
holds and the second order equation of motion is given by
\[
    \frac{d}{d \tau}
    \left(
        \frac{\partial L _{\mathrm{E    FF}}}{\partial \dot{R}}
    \right)
    -
    \frac{\partial L _{\mathrm{EFF}}}{\partial R}
    =
    0
    .
\]
Moreover, $H _{\mathrm{EFF}} \equiv 0$ is a constraint on the system. We also anticipate that
in this setup the expression for the \emph{Euclidean momentum}, i.e. the momentum along
a tunnelling trajectory, is
\begin{equation}
    P ^{\mathrm{(e)}} _{\mathrm{EFF}} (R , R')
    =
    R
    \left[
        \arctan
        \left(
            \frac{R'}{\epsilon \sqrt{f (R) - (R') ^{2}}}
        \right)
    \right]
    =
    - \imath P _{\mathrm{EFF}} (R , \dot{R})
    .
\label{eq:euceffmom}
\end{equation}
The last equality in the equation above suggests that the Euclidean system
can be obtained by Wick rotating the classical one\footnote{$R '$ denotes the derivative
of $R$ with respect to $\tau ^{(\mathrm{e})} = - \imath \tau$, so that $\dot{R} = \imath R '$.
As far as the spacetime structure is concerned, we also have to Wick rotate the time coordinates
in ${\mathcal{M}} _{\stPLUMIN}$, $t _{\stPLUMIN} ^{(\mathrm{e})} = - \imath t _{\stPLUMIN}$,
to obtain the corresponding Euclidean manifolds.}. This can be proved
deriving the Euclidean junction condition \cite{bib:NuPhy1990B339...417G}.

\section{\label{sec:genres}Some general results}

We now prove some general results to provide toeholds for the
discussion of the problems that will emerge in the Euclidean sector. Some of them
appear already in the literature, but not in a systematic
exposition. These results also provide simple, but useful consistency checks.
\begin{prop}
The junction condition and the effective potential are invariant under the relabelling
``$+ \leftrightarrow -$'', whereas the signs change as
$\epsilon _{\stPLUMIN} \rightarrow - \epsilon _{\stMINPLU}$.
\end{prop}
\proof{the validity of the above result for $V (R)$ and
$\epsilon _{\stPLUMIN}$ is manifest from (\ref{eq:claeffequ}) and (\ref{eq:epssig}). Then
the invariance of the junction condition (\ref{eq:juncon}) immediately follows.}
Another property directly related to the algebraic structure of (\ref{eq:juncon}), is the following.
\begin{prop}
\label{prop:f_RbigV_R}
The relations $V(R) \leq f _{\stPLUMIN} (R)$ always hold. Moreover, if $V(\bar{R}) = f _{\stPLUMIN} (\bar{R})$,
then $V (R)$ and $f _{\stPLUMIN} (R)$ are tangent at $R = \bar{R}$.
\end{prop}
\proof{the first result follows immediately after some algebra, since
\begin{equation}
    f _{\stPLUMIN} (R) - V (R)
    =
    \frac{\left( R ^{2} f _{\stMIN} - R ^{2} f _{\stPLU} \mp M ^{2} (R) \right) ^{2}}{4 M ^{2} (R) R ^{2}}
    =
    a _{\stPLUMIN} ^{2} (R)
    \geq 0
    .
\label{eq:radargsemposdef}
\end{equation}
This, together with the first derivative
of the equation above, $f ' _{\stPLUMIN} (R) - V ' (R) = 2 a (R) a ' (R)$ completes the proof of
the second result, because $f _{\stPLUMIN} (\bar{R}) = V (\bar{R})$, i.e. $a _{\stPLUMIN} (\bar{R}) = 0$,
implies $f ' _{\stPLUMIN} (\bar{R}) = V ' (\bar{R})$.}
Two immediate consequences then follow.
\begin{prop}
\label{prop:sigandtan}
Along a classical trajectory the signs $\epsilon _{\stPLUMIN}$ vanish at the points at which the potential $V (R)$
is tangent to the metric function $f _{\stPLUMIN} (R)$ and only at these points.
\end{prop}
\proof{the result follows from
(\ref{eq:epssig}) and (\ref{eq:radargsemposdef}), which show
that $\epsilon _{\stPLUMIN} = \mathrm{sign} (a _{\stPLUMIN} (R))$.}
\begin{prop}
\label{prop:radandsol}
The conditions about the well-definiteness of the two radicals in the junction condition
(\ref{eq:juncon}) do not impose any restriction on the classical/tunnelling solutions.
\label{prop:radno_res}
\end{prop}
\proof{from what we have seen above, if the effective (classical
looking) equation (\ref{eq:claeffequ}) is satisfied, the arguments of the radicals,
$\dot{R} ^{2} + f _{\stPLUMIN} (R) = f _{\stPLUMIN} (R) - V (R)$, are always nonnegative by
Prop.$\:$\ref{prop:f_RbigV_R}.}
Up to this point we have systematized some results concerning the mutual relationships
of the metric functions $f _{\stPLUMIN}$, the potential $V$ and the signs $\epsilon _{\stPLUMIN}$.
Some additional results are now obtained about the relative positions of the zeroes
of $\epsilon _{\stPLUMIN}$, $V$ and the positions of the zeroes of $f _{\stPLUMIN}$.
\begin{prop}
\label{prop:sigchanegf_R}
The signs $\epsilon _{\stPLUMIN}$ can change either i) along a classical solution of (\ref{eq:juncon}) in
a region in which $f _{\stPLUMIN} \leq 0$, or ii) along a tunnelling trajectory.
\end{prop}
\proof{let us assume that at the point $R = \bar{R}$
we have $\epsilon _{\stPLUMIN} (\bar{R}) = 0$; this implies $a _{\stPLUMIN} (\bar{R}) = 0$
by Prop.$\:$\ref{prop:sigandtan}. If we
are along a classical solution of (\ref{eq:juncon}) because of
(\ref{eq:claeffequ}) we must have $V (\bar{R}) \leq 0$, i.e. $f _{\stPLUMIN} (\bar{R}) \leq 0$. If instead
we have $f _{\stPLUMIN} (\bar{R}) > 0$, then we also have $V (\bar{R}) >0$, i.e. we are
along a tunnelling trajectory.}
An analogous result can then be proved for the turning points of the classical solutions.
\begin{prop}
\label{prop:turpoiposf_R}
A turning point $R _{0}$ of (\ref{eq:juncon}) always satisfies $f _{\stPLUMIN} (R _{0}) \geq 0$.
\end{prop}
\proof{we know that in general
$f _{\stPLUMIN} (R) - V (R) \geq 0$; thus at $R _{{\scriptscriptstyle{0}}}$, where $V(R _{{\scriptscriptstyle{0}}}) = 0$,
we have $f _{\stPLUMIN} (R _{{\scriptscriptstyle{0}}}) \geq 0$.}
Related to this result is the following one, which details the regularity properties of $P _{\mathrm{EFF}}$.
\begin{prop}
\label{prop:effmomdef}
The effective momentum $P _{\mathrm{EFF}}$
is always well defined on a classically allowed trajectory,
with the exception of the points at which $f _{\stPLUMIN}$ vanish.
\end{prop}
\proof{by Prop.$\:$\ref{prop:radno_res} the radicals that appear in the expression
of $P _{\mathrm{EFF}}$ are always well defined along a classical
solution of the junction condition. Critical points of $P _{\mathrm{EFF}}$ can thus only
appear either if they vanish or because of the presence of the inverse hyperbolic tangent,
whose argument must be in the interval $(-1,+1)$. Two complementary subcases can be singled out.\\
$f _{\stPLUMIN} \neq 0)$
Under this condition we can rule out the first possibility,
since along a classically allowed trajectory $\epsilon _{\stPLUMIN}$ can vanish only if $f _{\stPLUMIN} < 0$,
when the radical is in the numerator. Troubles with the inverse hy\-per\-bo\-lic tangent are
quickly excluded as well. If $f _{\stPLUMIN} > 0$, then the exponent of its
argument is $+1$, and we can forget about it. At the same time the absolute value of the numerator
is lower than the one of the denominator, so the momentum is well defined in this case.
If $f _{\stPLUMIN} < 0$, then the absolute value of the numerator \emph{is} bigger than the
one of the denominator, but the ratio is raised to the power $-1$ and
still there is no problem.\\
$f _{\stPLUMIN} = 0)$
This case is more subtle; if the corresponding $\epsilon _{\stPLUMIN}$ sign is non vanishing,
then the absolute value of the argument of the inverse hyperbolic tangent is equal to $1$ and
the momentum has a logarithmic, i.e. integrable, divergence. If, instead, also $\epsilon _{\stPLUMIN} = 0$,
a \emph{case by case} analysis is required.\\
\noindent{}We can, thus, conclude that the momentum can be \emph{not well defined} only at the points
where $f _{\stPLUMIN} = 0$; if $\epsilon _{\stPLUMIN} \neq 0$ it has a logarithmic divergence;
a \emph{case by case} analysis has instead to be done if $\epsilon _{\stPLUMIN} = 0$.}
The regularity of the Euclidean effective momentum can be analyzed in general as well.
\begin{prop}
\label{prop:euceffmomdef}
The Euclidean effective momentum $P ^{(\mathrm{e})} _{\mathrm{EFF}}$
is always well defined along a tunnelling trajectory in the sense that, at most,
it can have discontinuities.
\end{prop}
\proof{from a quick inspection of
(\ref{eq:euceffmom}) and from the result in Prop.$\:$\ref{prop:radno_res}
we see that the only trouble to the momentum can come from vanishing radicals
in the denominator, i.e. from vanishing $\epsilon _{\stPLUMIN}$. This can
indeed happen along a tunnelling trajectory (Prop.$\:$\ref{prop:sigchanegf_R})
and if this happens \emph{inside} it, e.g. at $\bar{R}$, the argument of the corresponding inverse tangent function
tends to $\pm \infty$ there. If $\lim _{R \to \bar{R} ^{+}} \neq \lim _{R \to \bar{R} ^{-}}$
in the standard branch of the $\arctan$ function a discontinuity appears. This discontinuity
can be eliminated by choosing another appropriate branch, although this will, in general,
affect the value of $P ^{(\mathrm{e})} _{\mathrm{EFF}}$ at, at least, one turning point.
If $\bar{R}$ coincides with a turning point, i.e. it occurs not \emph{inside} but \emph{at the boundary}
of a tunnelling trajectory, a \emph{case by case} analysis is again required.}

The physical content of these results can be expressed in general terms using the
concepts of $R ^{\stPLUMIN}$ and $T ^{\stPLUMIN}$ introduced before. We  have, in fact, proved that:
\begin{enumerate}
\item the $\epsilon _{\stPLUMIN}$ signs can only vanish either i) in the closure of a $T ^{\stPLUMIN}$ region along a classical
solution of the junction condition (\ref{eq:juncon}) or ii)  along a tunnelling trajectory;
\item turning points can only be present in the closure of an $R ^{\stPLUMIN}$ region; this second fact is
coherent with the fact that in the Euclidean description of the manifolds ${\mathcal{M}} _{\stPLUMIN}$,
there is no region corresponding to the $T ^{\stPLUMIN}$ ones; loosely speaking, the shell has
\emph{nowhere} to tunnel from one of these regions!
\end{enumerate}

These results are valid for arbitrary spherically symmetric junctions, independently from the matter
content of spacetime and/or of the shell. It is also noteworthy to stress the following
particular case, which happens when a turning point is exactly on the boundary of an $R$-region of one of
the two spacetimes (i.e. at a point where at least one of $f _{\stPLUMIN}$ vanishes). In this
case one of the $f _{\stPLUMIN}$ is zero, but $V$ is also zero, so the corresponding $\epsilon _{\stPLUMIN}$
sign is zero too. We call this the \emph{exceptional case}, but we will not consider it further here.

\section{\label{sec:tunpro}Tunnelling problems}

The above results can be used to obtain general insight non only about the classical dynamics of the system,
but, especially, about the semiclassical one. Here we will not discuss the interesting possibilities of
using WKB methods to determine the semiclassical stationary states
\cite{bib:SemicState,bib:ClQuG2002..19..6321A}, but we
will, instead, concentrate on the tunnelling process. In particular, it is possible to use
the Euclidean momentum (\ref{eq:euceffmom}) to calculate the value of the tunnelling action
(i.e of the probability)
\[
    S _{\mathrm{TUN}} = \int _{R _{1}} ^{R _{2}} P _{\mathrm{EFF}} ^{\mathrm{(e)}} (R) d R
    \quad \mathrm{where} \quad
    P _{\mathrm{EFF}} ^{\mathrm{(e)}} (R) = \left . P _{\mathrm{EFF}} ^{\mathrm{(e)}} (R , R') \right | _{R ' = \sqrt{V (R)}}
    ;
\]
$P _{\mathrm{EFF}} ^{\mathrm{(e)}} (R)$ is the Euclidean Momentum \emph{evaluated along a tunnelling trajectory}. This
has been done for various configurations and results in agreement with other independent calculations have been obtained
(for instance the results by Coleman and de Luccia \cite{bib:PhReD1980..21..3305L} and by Parke \cite{bib:PhLeB1983.121...313P},
can be reproduced). Unfortunately things do not always work out
so smoothly: problems arise when at least one of the $\epsilon _{\stPLUMIN}$ signs vanishes along the tunnelling
trajectory. These problems relate i) to the ones connected with the difficulty to build the Euclidean
manifold interpolating between the classical spacetime configuration described by the pre- and post-tunnelling
solutions of the junction condition \cite{bib:NuPhy1990B339...417G}
and ii) to the ones connected with the difference in the tunnelling description given
by canonical and path-integral methods \cite{bib:NuPhy1990B339...417G}.
It is convenient to summarize these issues using as a definite model,
the case in which we have a de Sitter/Schwarzschild junction
( $f _{\stPLU} (r _{\stPLU}) = 1 - \chi ^{2} r _{\stPLU} ^{2}$,
$f _{\stMIN} (r _{\stMIN}) = 1 - 2 m / r _{\stMIN}$) by a matter shell with equation of state $p = - \sigma$, $\sigma$
being the tension of the shell (i.e. $M (R) = 4 \pi \sigma R ^{2}$).

\begin{figure}[htb]
\begin{center}
\rule{\textwidth}{0.5pt}\\[3mm]
\begin{tabular}{|c|c|c|}
\hline
\includegraphics[width=4.5cm]{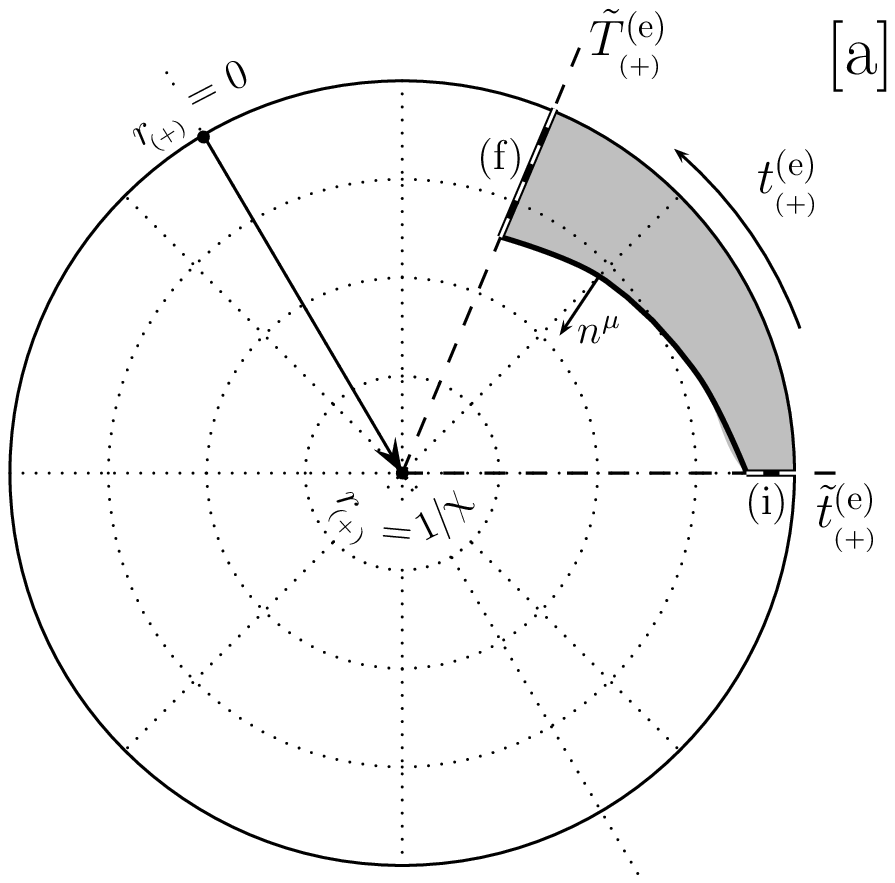}
&
\includegraphics[width=4.5cm]{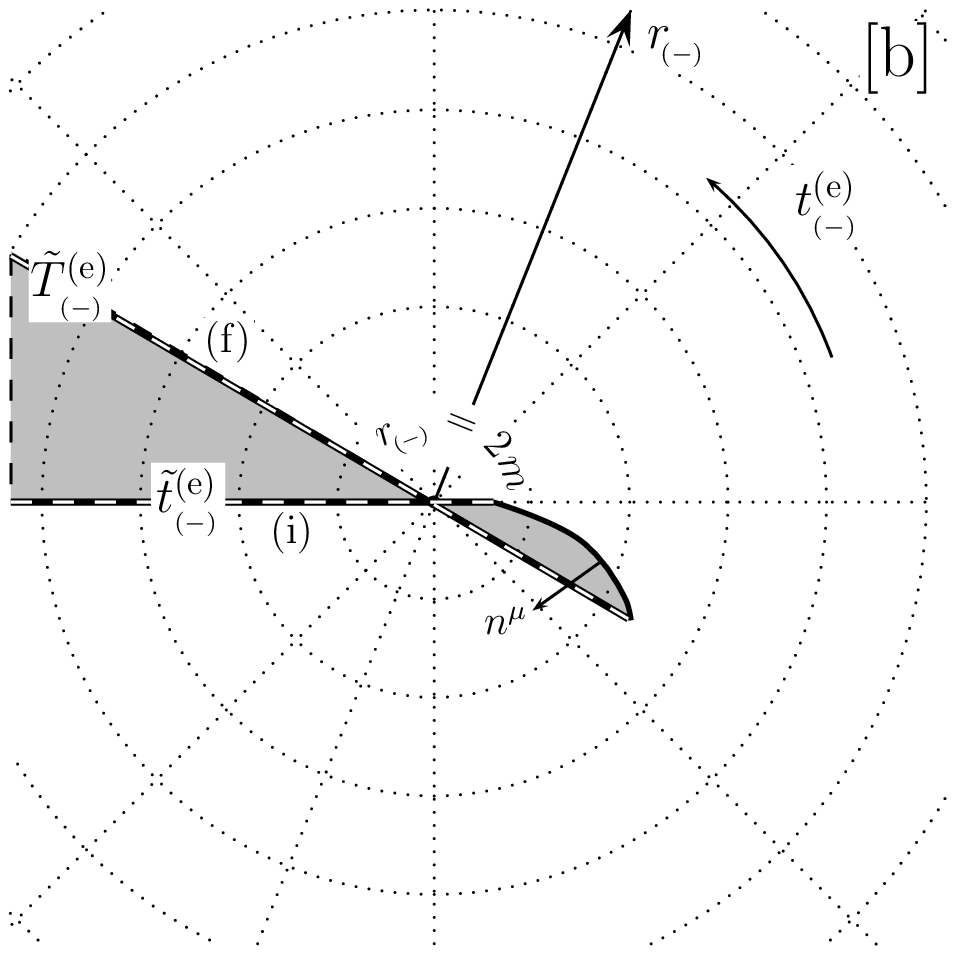}
&
\includegraphics[width=4.5cm]{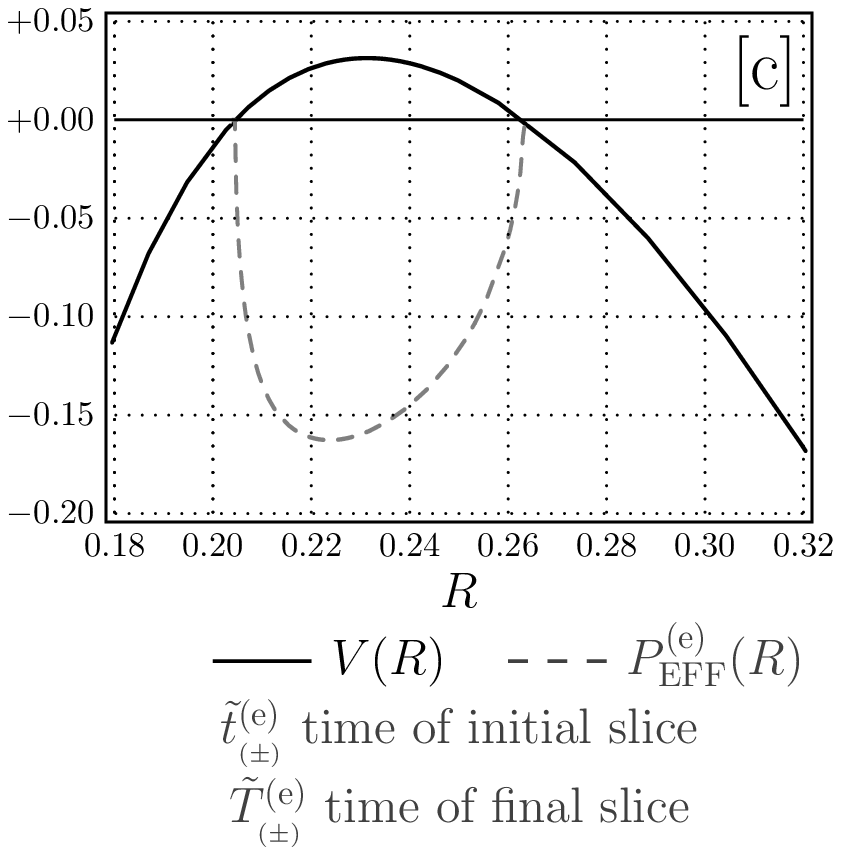}
\\
\hline
\end{tabular}
\caption{\label{fig:firtuncas}{\footnotesize{}A case of tunnelling where no problems appear: in [a] and [b] the
Euclidean parts of spacetime participating in the junction are the grayed areas. Initial and final slices
are labelled (i) and (f) respectively. The Euclidean trajectory of the shell is the solid curve.
Initial and final Euclidean times and the normal are also shown.}}
\end{center}
\end{figure}

To this end, in Fig.$\:$\ref{fig:firtuncas} we first analyze a situation free from troubles.
In panels [a] and [b] we see
that the normal is always transverse to the constant $r _{\stPLUMIN}$ surfaces. According
to the definitions of $\epsilon _{\stPLUMIN}$, they then do not vanish, so that \emph{via}
Prop.$\:${\protect\ref{prop:euceffmomdef}} we know that the momentum does not have any troubles along the tunnelling
trajectory, as shown in panel [c].

\begin{figure}[htb]
\begin{center}
\begin{tabular}{|c|c|c|}
\hline
\includegraphics[width=4.5cm]{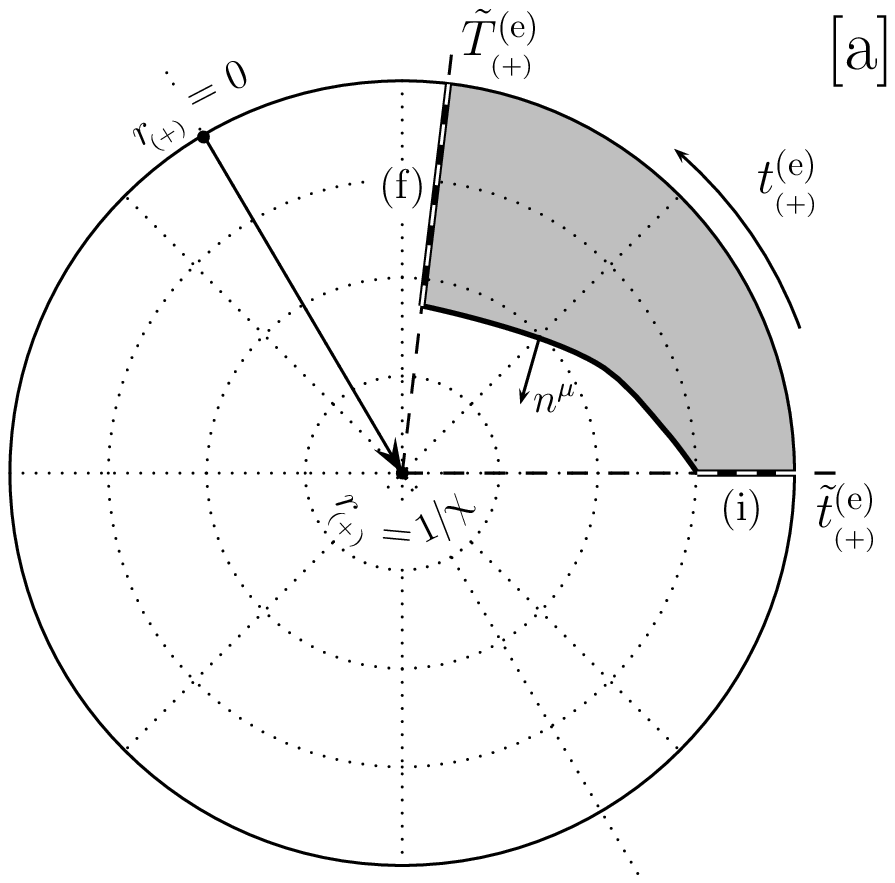}
&
\includegraphics[width=4.5cm]{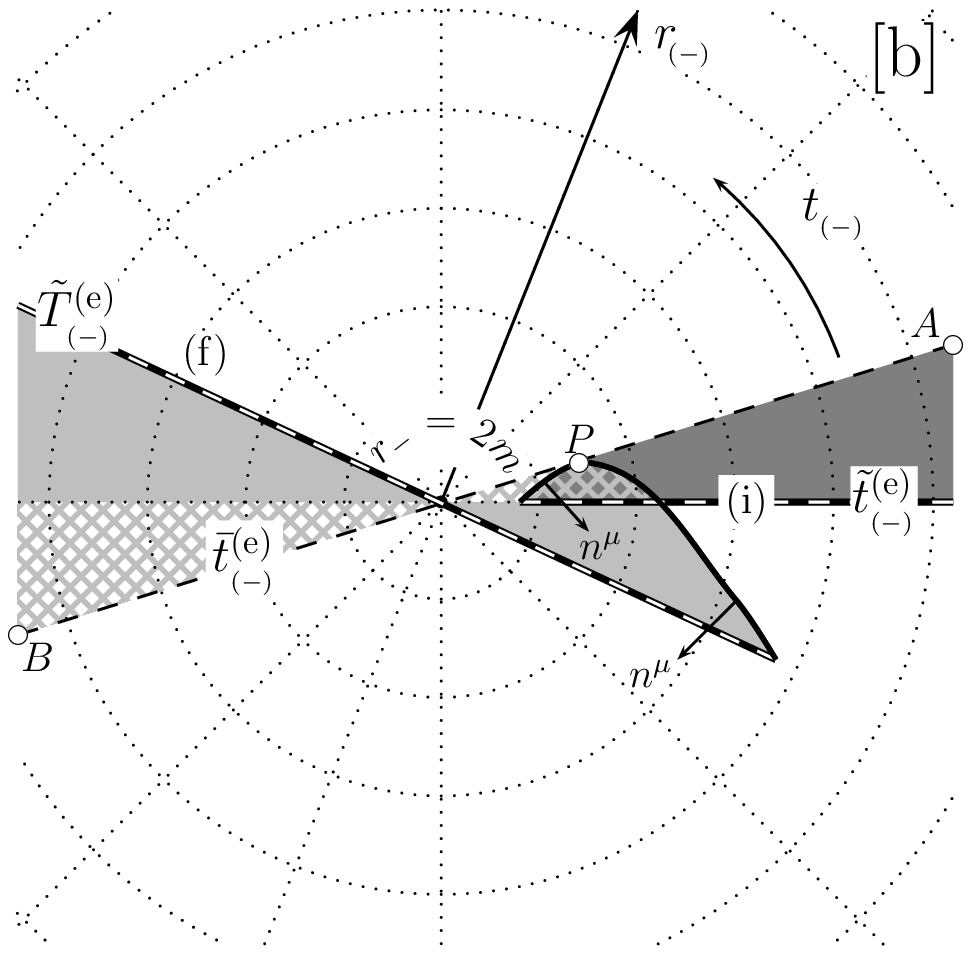}
&
\includegraphics[width=4.5cm]{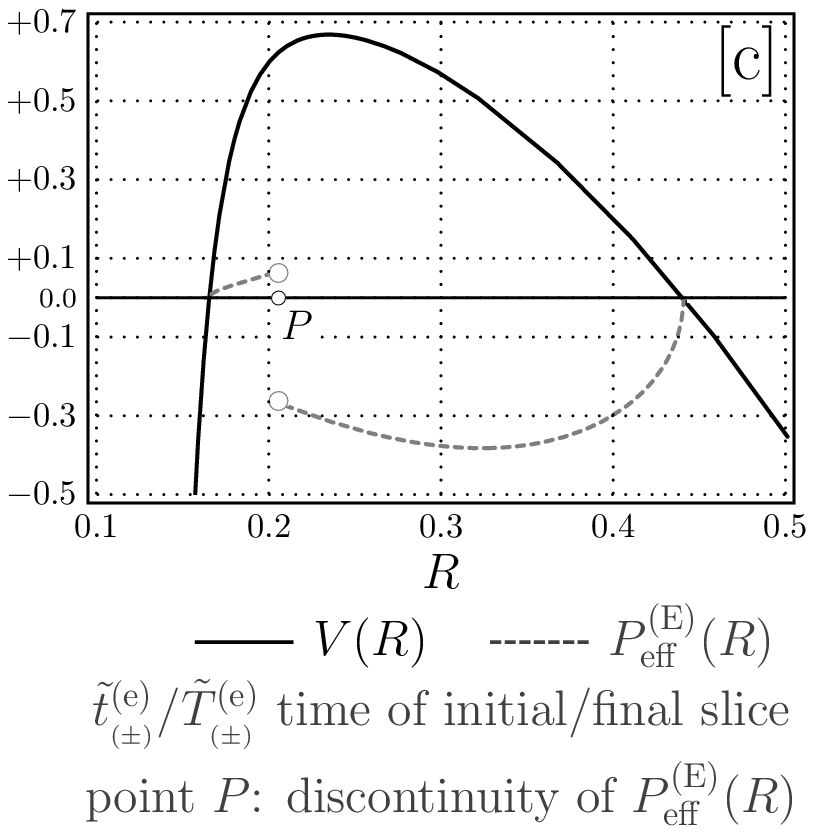}
\\
\hline
\end{tabular}
\caption{\label{fig:sectuncas}{\footnotesize{}A case of tunnelling when one part of the spacetime
participating in the junction is problematic. If we proceed as in the case of Fig. {\protect\ref{fig:firtuncas}}
we encounter soon a point $P$ after which, the determination of the part of spacetime participating in
the junction becomes unclear (see the text for details).}}
\rule{\textwidth}{0.5pt}
\end{center}
\end{figure}

A different situation is, instead, shown in Fig.$\:$\ref{fig:sectuncas}.

The de Sitter part of the junction in panel [a] is fine as before, but the Schwarzschild one in panel
[b] has a peculiar feature: there is a point $P$ along the trajectory at which the normal is not
transverse to the constant $r _{\stMIN}$ surface, so that $\epsilon _{\stMIN} (P) = 0$. Thus, we
have some difficulty in identifying the part of the Schwarzschild spacetime participating in the
junction, since the small area which is both \emph{dark-grayed} and \emph{crossed-hatched} is covered twice by the
evolution of the Euclidean spacetime slice. This difficulty in identifying the instanton is reflect
by the momentum plot in panel [c]: a discontinuity appears, as expected from
Prop.$\:${\protect\ref{prop:euceffmomdef}}. Notice that if we naively take the union of the grayed
regions as the part of spacetime participating in the junction, a strange boundary (the $AB$ line)
appears\footnote{There are reasonable proposals {\protect\cite{bib:NuPhy1990B339...417G}}
to deal with this problem, but we will not discuss
them here. The point we would like to stress, focusing on the problem rather than the possible solutions,
is that particular care must be used when dealing with the Euclidean junction.}.

\begin{figure}[htb]
\begin{center}
\rule{\textwidth}{0.5pt}\\[3mm]
\begin{tabular}{|c|c|c|}
\hline
\includegraphics[width=4.5cm]{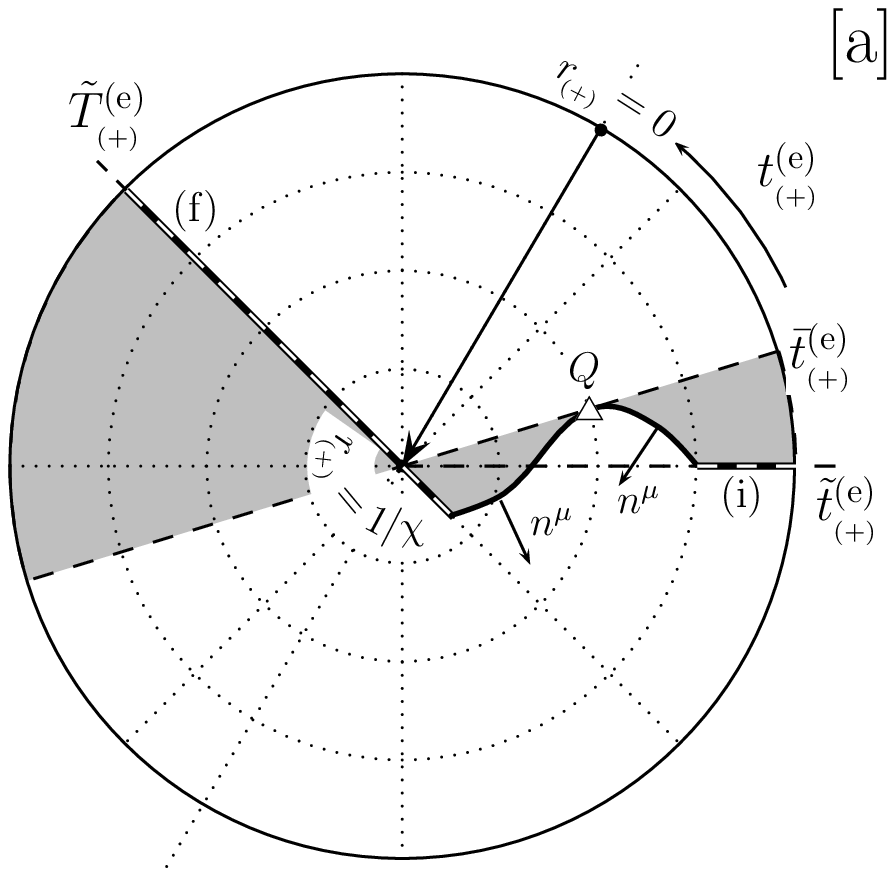}
&
\includegraphics[width=4.5cm]{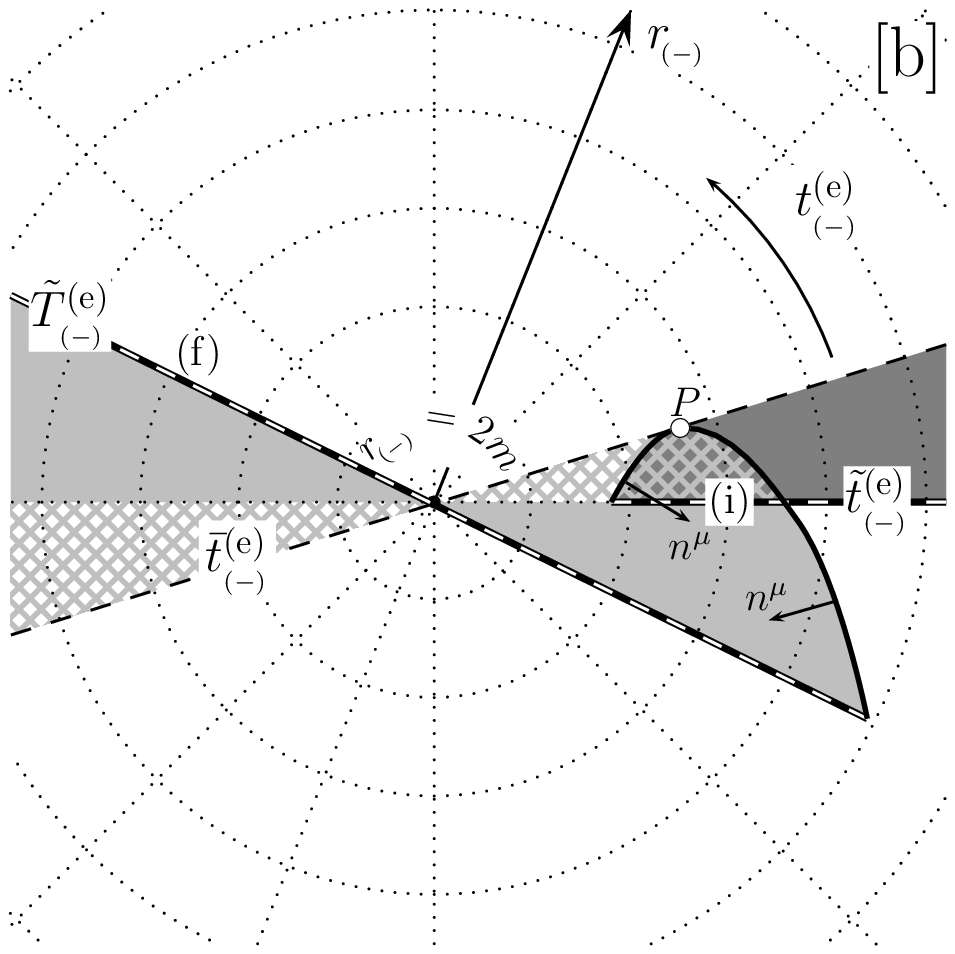}
&
\includegraphics[width=4.5cm]{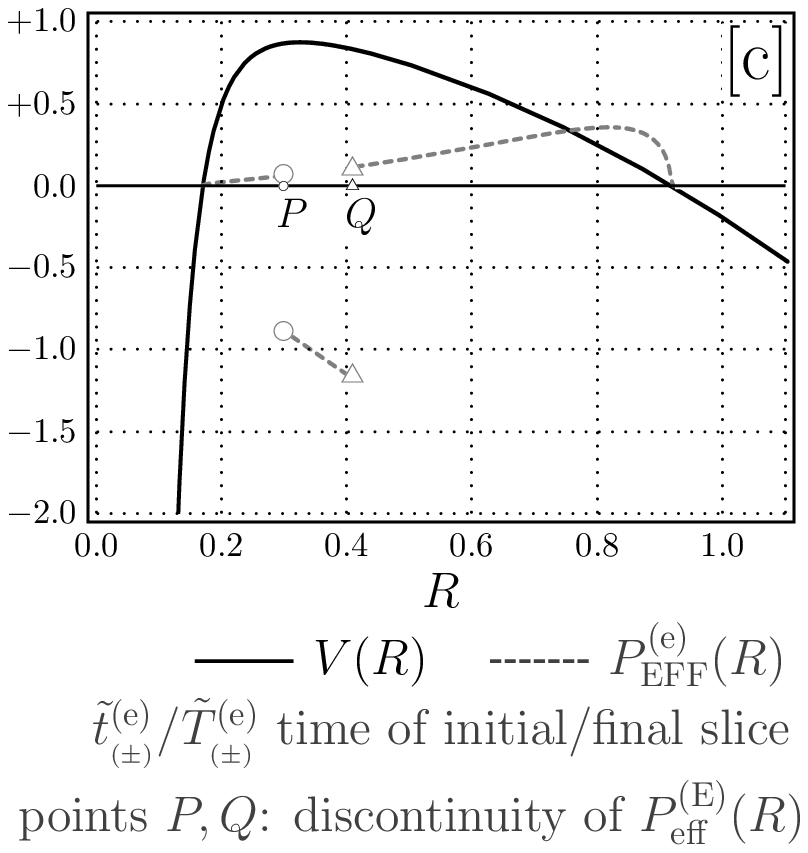}
\\
\hline
\end{tabular}
\caption{\label{fig:thituncas}{\footnotesize{}The problem which appeared in Fig.$\:${\protect\ref{fig:sectuncas}}
is generic and can affect both spacetimes, separately, or at once, as in this case. Now, in panel
[a] some problems appear in connection with the existence of point $Q$, whereas in panel [b] the
situation maintains the same difficulty encountered in Fig.$\:${\protect\ref{fig:sectuncas}}. Two discontinuities
of $P _{\mathrm{EFF}} ^{(\mathrm{e})} (R)$ are present (panel [c]).}}
\end{center}
\end{figure}

The problem shown above is not typical of the Schwarzschild patch and can appear also in
the de Sitter one, as shown in Fig.$\:$\ref{fig:thituncas}.

In both diagrams, at $Q$ in panel [a] and at $P$ in panel [b], the normal to the shell
become non-transverse to the constant $r _{\stPLUMIN}$ surfaces.
At these points the corresponding signs $\epsilon _{\stPLUMIN}$ vanish and the momentum
develops a discontinuity, as shown in panel [c]. It is non trivial to build the junction and
we face again a double covering with inverted normal direction in the Schwarzschild patch
during the evolution of the initial slice (i) into the final one (f).

Concerning the last two cases, we notice, as we did at the end of the proof of
Prop.$\:$\ref{prop:euceffmomdef}, that, in view of the form of the effective momentum
$P _{\mathrm{EFF}} ^{\mathrm{(e)}}$, it is certainly possible to choose appropriate branches of the
inverse tangent functions to \emph{cure} the discontinuity:
nevertheless, this spoils the vanishing of the momentum at one of the turning points. It would,
moreover, be interesting, to understand these possibilities in connection with the structure
of the Euclidean spacetime diagram.

\section{\label{sec:dissec}Discussion and Conclusion}

Remembering the results presented in Sec.$\:$\ref{sec:genres}, we would like to point out
that the problems discussed in the previous section under very specific settings,
are in fact general ones. In our opinion this fact has not yet been properly recognized.
We would also like to make clear that, although the problems with the effective momentum
\emph{could} be solved by arguing that it is the effective theory that has intrinsic limits,
\emph{we think that this partial solution would be rather unsatisfactory}.
The main point we are trying to make here is that the problems of the effective
formulation are \emph{closely tied} to the geometric properties of the Euclidean solution
to the junction condition and to the \emph{Euclidean structure of the spacetime patches}
that participate in the junction. Moreover, the difficulties described above
\emph{can} be absent, and \emph{when they are absent}, perfectly consistent results
are obtained. In this respect, it becomes even more suggestive to draw the following picture.
The idea of bubbles/shells tunnelling has been originally developed to overcome some
weak points in the description provided by purely classical models of vacuum bubbles using
general relativistic shells: in particular, the problem of initial singularity, i.e. the fact
that exponentially expanding solutions giving rise to a baby universe, classically, have a
singularity in their past. Tunnelling (i.e. the use of quantum effects) solves this issue:
in fact, we can start with a solution regular in the past (the bounded solution, which
classically would never grow enough) and have it tunnel into an infinitely expanding one
(the unbounded/bounce solution) which has the late time behavior we are interested in,
but cannot exist at early times. We are, thus \emph{using quantum effects to circumvent
the consequences of classical singularity theorems}. The same idea has been employed also
in a logically opposite direction using different spacetimes to build up the junction:
then the reversed tunnelling process can describe a collapsing shell of matter
\cite{bib:ShellCollapse},
which classically doomed to crash in a future singularity, is, instead, saved by
quantum effects, again avoiding the fate prescribed by classical singularity theorems.
Another fact which has not yet been appreciated is that, in view of the simple
but general results of Sec.$\:$\ref{sec:genres}, this kind of processes is affected by
the same difficulties, again related to a hard to interpret behavior in the Euclidean
sector. In this sense the more general formulation of the problems
shown in Figs.$\:$\ref{fig:firtuncas} and \ref{fig:sectuncas}, that can be
obtained in terms of the results of Sec.$\:$\ref{sec:genres}, shows that they are
\emph{very general issues which appear when we try to use semiclassical quantum
effects to circumvent the consequences of classical singularity theorems}. As we said,
this issues are not restricted to applications to the cosmological (baby-universes)
scenario, as it is often believed, but represent instead another manifestation
of the intrinsic difficulty in the interplay between the properties
of general relativity and those of quantum theory. It seems to us
a great opportunity, given to us by the intuitive and beautiful geometric character
of Israel junction conditions, that these issues can appear at a technically
rather simple level, offering us the possibility
to concentrate our attention on their physical/geometric significance.
This study is
still work in progress, which at present is focused on finding general geometric
criteria i) to identify the appearance of the above issues in the junction
tunnelling process and ii) to characterize the difficulties in the Euclidean sector in terms
of intuitive properties of the effective theory. Additional results will be reported
(hopefully soon) elsewhere.
\paragraph*{Acknowledgements.}\hspace{-5mm}
\noindent{\small{}The author would like to thank A. Guth for some stimulating
discussion on the subject of this work, which is partly supported by an Invitation
Fellowship of the \emph{Japan Society for the Promotion of Science} and has been partly
supported by a grant of the \emph{Fulbright Commission}.}


\begin{thebibliography}{99}%
{\footnotesize{}%
\bibitem{bib:VacuumDecay}
S. {Coleman},
{\em Phys. Rev. D}, \textbf{15} 2929, 1977;
Jr. Curtis~G. {Callan} and S. {Coleman},
{\em Phys. Rev. D}, \textbf{16} 1762, 1977.
\bibitem{bib:PhReD1980..21..3305L}
S. {Coleman} and F.~De {Luccia},
{\em Phys. Rev. D}, \textbf{21} 3305, 1980.
\bibitem{bib:Bubbles01}
K. {Sato}, M. {Sasaki}, H. {Kodama}, and K.-I. {Maeda},
{\em Progr. Theor. Phys.} \textbf{65} 1443, 1981;
H. {Kodama}, M. {Sasaki}, K. {Sato}, and K.-I. {Maeda},
{\em Progr. Theor. Phys.} \textbf{66} 2052, 1981;
K. {Sato},
{\em Progr. Theor. Phys.} \textbf{66} 2287, 1981;
K.-I. {Maeda}, K. {Sato}, M. {Sasaki}, and H. {Kodama},
{\em Phys. Lett. B} \textbf{108} 98, 1982;
K. {Sato}, H. {Kodama}, M. {Sasaki}, and K.-I. {Maeda},
{\em Phys. Lett. B} \textbf{108} 103, 1982;
H. {Kodama}, M. {Sasaki}, and K. {Sato},
{\em Progr. Theor. Phys.} \textbf{68} 1979, 1982;
S.~W. Hawking, I.~G. Moss, and J.~M. Stewart,
{\em Phys. Rev. D} \textbf{26} 2681, 1982;
W.~Z. Chao,
{\em Phys. Rev. D} \textbf{28} 1898, 1983.
\bibitem{bib:PhReD1987..35..1747G}
S.~K. {Blau}, E.~I. {Guendelman}, and A.~H. {Guth},
{\em Phys. Rev. D} \textbf{35} 1747, 1987.
\bibitem{bib:Bubbles02}
V.~A. {Berezin}, V.~A. {Kuzmin}, and I.~I. {Tkachev},
{\em Phys. Rev. D} \textbf{36} 2919, 1987;
A.~Aurilia, R.~S. Kissack, R.~Mann, and E.~Spallucci,
{\em Phys. Rev. D} \textbf{35} 2961, 1987;
K. {Lee} and E.~J. {Weinberg},
{\em Phys. Rev. D} \textbf{36} 1088, 1987.
\bibitem{bib:NuPhy1990B339...417G}
E. {Farhi}, A.~H. {Guth}, and J. {Guven},
{\em Nucl. Phys.} \textbf{B339} 417, 1990.
\bibitem{bib:QuantumModels}
W. {Fishler}, D. {Morgan}, and J. {Polchinski},
{\em Phys. Rev. D} \textbf{41} 2638, 1990;
---{\em ibid.} \textbf{42} 4042, 1990;
V.~A. {Berezin}, V.~A. {Kuzmin}, and I.~I. {Tkachev},
{\em Phys. Rev. D} \textbf{43} R3112, 1991;
M.~Sasaki, T.~Tanaka, K.~Yamamoto, and J.~Yokoyama,
{\em Phys. Lett.} \textbf{B317} 510, 1993;
T.~Tanaka,
{\em Nucl. Phys.} \textbf{B556} 373, 1999;
A.~Khvedelidze, G.~V. Lavrelashvili, and T.~Tanaka,
{\em Phys. Rev. D} \textbf{62} 083501, 2000.
\bibitem{bib:QuantumTroubles}
S. Ansoldi and L. Sindoni,
in the {\em Proceedings of the 6th International Symposium
on Frontiers of Fundamental Physics} (FFP6, Udine, Italy, 2004),
p. 69, eprint: \texttt{gr-qc/0411042};
S. Ansoldi,
in the {\em Proceedings of the 16th Workshop on General Relativity
and Gravitation} (JGRG16, Niigata, Japan, 2006),
{\em K.-I. Oohara, T. Shiromizu, K.-I. Maeda aand M. Sasaki editors},
p. 114, eprint: \texttt{gr-qc/0701117};
A. {Aguirre} and M.~C {Johnson},
{\em Phys. Rev. D} \textbf{72} 103525, 2005.
\bibitem{bib:ClQuG2002..19..6321A}
S. {Ansoldi},
{\em Class. Quantum Grav.} \textbf{19} 6321.
\bibitem{bib:ExtraBiblio}
S.~{Ansoldi} and L.~{Sindoni},
{\em Phys. Rev. D} in print, September 2007,
eprint: \texttt{arXiv:0704.1073 [gr-qc]}.
\bibitem{bib:Freem1970...1..1279W}
C.~W. {Misner}, K.~S. {Thorne}, and J.~A. {Wheeler},
{\em "Gravitation"}
(W.~H. Freeman and Company, San Francisco, 1970).
\bibitem{bib:PhReD1989..40..2511S}
A. {Aurilia}, M. {Palmer}, and E. {Spallucci},
{\em Phys. Rev. D} \textbf{40} 2511, 1989.
\bibitem{bib:ClQuG1997..14..2727S}
S. {Ansoldi}, A. {Aurilia}, R. {Balbinot}, and E. {Spallucci},
{\em Class. Quantum Grav.} \textbf{14} 2727, 1997.
\bibitem{bib:IsraelJunctions}
W. {Israel},
{\em Nuovo Cimento}, \textbf{B44} 1, 1966;
---{\em errata ibid.} \textbf{B48} 463, 1967;
C. {Barrabes} and W. {Israel},
{\em Phys. Rev. D} \textbf{43} 1129, 1991.
\bibitem{bib:TRRegions}
I.~D. Novikov,
{\em Commun. Sternberg Astron. Inst.} \textbf{132} 3, 1964;
---{\em ibid.} 43, 1964.
\bibitem{bib:ClQuG2001..18..2195O}
V. {Berezin} and M. {Okhrimenko},
{\em Class. Quantum Grav.} \textbf{18} 2195, 2001.
\bibitem{bib:LagFormShells}
P.~{Hajicek} and J.~{Bicak},
{\em Phys. Rev. D} \textbf{56} 4706, 1997;
J.~L. {Friedman}, J. {Louko}, and S.~N. {Winters-Hilt}.
{\em Phys. Rev. D} \textbf{56}  7674, 1997;
P. {Hajicek} and J. {Kijowski},
{\em Phys. Rev. D} \textbf{57} 914, 1998;
---{\em errata ibid.} \textbf{61} 129901, 2000;
P. {Hajicek},
{\em Phys. Rev. D} \textbf{57} 936, 1998;
J.~L. {Friedman}, J. {Louko}, and B.~F. {Whiting},
{\em Phys. Rev. D} \textbf{57} 2279, 1998;
P. {Hajicek},
{\em Phys. Rev. D} \textbf{58} 084005, 1998;
S. {Mukohyama},
{\em Phys. Rev. D} \textbf{65} 024028, 2002;
R. {Capovilla}, J. {Guven}, and E. {Rojas},
{\em Class. Quantum Grav.} \textbf{21} 5563, 2004;
C. Barrabes and W. Israel;
{\em Phys. Rev. D} \textbf{71} 064008, 2005.
\bibitem{bib:SemicState}
M. {Visser},
{\em Phys. Rev. D} \textbf{43} 402, 1991;
V.~A. {Berezin},
{\em Phys. Lett. B} \textbf{241} 194, 1990;
S. Ansoldi,
{\em AIP Conf. Proc.} \textbf{751} 159, 2005;
S. Ansoldi,
in the {\em Proceedings of the 11th Marcell Grossman Meeting on General
Relativity} (Berlin, Germany, 2006) eprint: \texttt{gr-qc/0701082}.
\bibitem{bib:PhLeB1983.121...313P}
S. Parke,
{\em Phys. Lett.}, \textbf{121B} 313, 1983.
\bibitem{bib:ShellCollapse}
V. {Berezin},
{\em Int. J. Mod. Phys. D} \textbf{5} 679, 1996;
P. {Hajicek}, S. {Kay}, and K.~V. {Kuchar},
{\em Phys. Rev. D} \textbf{46} 5439, 1992;
K.~V. Kuchar,
{\em Phys. Rev. D} \textbf{50} 3961, 1994;
V. {Berezin},
{\em Phys. Rev. D} \textbf{55} 2139, 1997;
V. {Berezin}.
{\em Int. J. Mod. Phys. A} \textbf{17} 979, 2002.}
\end{thebibliography}
\end{document}